\documentclass[12pt,aps,manuscript]{paper}
\usepackage[T1]{fontenc}
\usepackage[latin9]{inputenc}
\usepackage[letterpaper]{geometry}
\usepackage{color}
\usepackage{amsmath}
\usepackage{amssymb}
\usepackage{graphicx}
\usepackage{setspace}
\usepackage{esint}
\usepackage[unicode=true,pdfusetitle,
 bookmarks=true,bookmarksnumbered=false,bookmarksopen=false,
 breaklinks=false,pdfborder={0 0 1},backref=false,colorlinks=true]
 {hyperref}
\hypersetup{
 linktocpage, citecolor=blue,linkcolor=blue,urlcolor=blue}

\makeatletter

\numberwithin{equation}{section}
\newcommand{\lyxaddress}[1]{
\par {\raggedright #1
\vspace{1.4em}
\noindent\par}
}

\newcommand{\Lbarsquare}{\bar{l}_s^2}
\newcommand{\Xvec}{\mathbf{X}}

\makeatother

\begin{document}

\title{Heavy-Quark Potential from Gauge/Gravity Duality: A large $D$ Analysis}

\author{Vikram Vyas}

\maketitle

\lyxaddress{Department of Physics, Shiv Nadar University, \\
 Gautam Budh Nagar, 203207, UP, India.\\
and\\
St. Stephen's College, Delhi University, Delhi 110007, India\\
 Email: \emph{vikram.phy.qcd@gmail.com}%
\footnote{On leave of absence from St. Stephen's College, Delhi University,
Delhi 110007, India%
}}
\begin{abstract}
The heavy-quark potential is calculated in the framework of gauge/gravity
duality using the large-$D$ approximation, where $D$ is the number
of dimensions transverse to the flux tube connecting a quark and an
antiquark in a flat $D+2$-dimensional spacetime. We find that in
the large-$D$ limit the leading correction to the ground-state energy,
as given by an effective Nambu-Goto string, arises not from the heavy
modes but from the behavior of the massless modes in the vicinity
of the quark and the antiquark. We estimate this correction and find
that it should be visible in the near-future lattice QCD calculations
of the heavy-quark potential.
\end{abstract}
\newpage{}

\tableofcontents{}

\section{Introduction}

There is strong numerical evidence that the dynamics of a flux tube
formed between a heavy-quark and an antiquark is well described by
the Nambu-Goto string in four spacetime dimensions (see, e.g., \cite{HariDass:2006pq,Brandt:2009tc,Brandt:2010bw},
and \cite{Teper:2009uq} for a review of the effective string approach
to QCD flux tubes with a review of the results for the closed flux
tubes). In particular, the ground-state energy of such a flux tube
matches very well with the Arvis formula for the ground-state energy
of an open Nambu-Goto string with fixed end points in four dimensions
\cite{Arvis:1983fp}. Yet one expects on general grounds that there
should be corrections to Arvis formula, for after all, the QCD flux
tube is not a one-dimensional object and has some ``intrinsic''
thickness that we expect to be related to the \emph{finite }correlation
length of the confining $\mathrm{SU(3)}$ gauge theory \cite{Wilson:kx}.
If one thinks of the fluctuations in the intrinsic thickness as an
additional degree of freedom, then integrating out this degree of
freedom would lead to an effective string action. Such an action,
in general, should contain all possible terms consistent with reparameterization
invariance. In the spirit of effective field theories, these terms
can be organized into relevant, marginal and irrelevant terms (for
a review of these ideas see, for e.g, Refs. \cite{David:1987uq,Dubovsky:2012uq,Aharony:2013fk}.)
The Nambu-Goto action is the only relevant term; then there is a marginal
term--the so-called extrinsic-curvature term--and infinitely many
irrelevant terms corresponding to higher and higher derivatives of
the string coordinates. Further, since we are considering open strings,
there could also be boundary terms. In fact, we know on very general
grounds that the resulting effective string action cannot be just
the Nambu-Goto action \cite{Polchinski:1991ax,Gliozzi:2012fk}.

To obtain an effective string action for QCD flux tubes--starting
from the fundamental description in terms of quarks and gluons--we
need to integrate out all the fluctuations of the gauge fields whose
wavelength is less than the correlation length of the confining theory.
This task is as difficult as deriving an effective chiral Lagrangian
for pions starting from the QCD action. The gauge/gravity duality
\cite{Aharony:2000aa}, inspired by the AdS/CFT correspondence \cite{Maldacena:1997re,Gubser:1998bc,Witten:1998qj,Witten:1998zw},
provides us with an alternate--though approximate and often heuristic--way
of analyzing the dynamics of strongly coupled gauge theories. Specifically,
it provides us with a geometrical way of exploring the consequences
of one of the defining properties of $\mathrm{SU(N)}$ gauge theories,
namely the existence of a finite correlation length of the gauge fields.
According to the gauge/gravity duality, a QCD flux tube should be
thought of as a holographic projection of a fundamental string in
five-dimensional curved space (that may have more compact directions
but they will not play a role in the present investigation.) This
picture then suggests that for confining gauge theories the intrinsic
thickness of the flux tube can be related to the position of a fundamental
string in the fifth dimension \cite{Danielsson:1998wt,Polchinski:2001ju}.
It was also pointed out in Ref. \cite{Vyas:2010uq} that this relationship
was consistent with the expected behavior of the intrinsic thickness
of a Nielsen-Olesen-like flux tube with its string tension \cite{Nielsen:1973cs}.

The aim of the present investigation is to use the geometrical picture
provided by the gauge/gravity duality to delineate the various sources
of correction to the heavy-quark potential as compared to the ground-state
energy of a Nambu-Goto string in flat four-dimensional spacetime.
These corrections are induced by the fact that the fundamental QCD
string lives in a five-dimensional curved space. We will analyze this
situation using the large-$D$ approximation, where $D+2$ is the
number of dimensions of the flat spacetime where the quark lives.
$D$ is therefore the number of directions transverse to a QCD flux
tube. The main advantage that this approach offers is that it treats
the massless modes of the string nonperturbatively.

In gauge/gravity duality, the confining property of the boundary gauge
theory is reflected in the fact that a fundamental string in five
dimensions can minimize its energy by staying at a fixed value of
its fifth coordinate as shown in Fig. \ref{fig:Minimal-Surface}.
We first calculate the large-$D$ behavior of the fundamental string
in the approximation where the fundamental string descends from its
minimal energy value to the boundary right at $x=0$ and $ $$x=L$,
where the quark and the antiquark are placed. This is shown in Fig.
\ref{fig:approxMinSurface}. The dynamics of the open string in this
approximation is very much like that of an open string placed at a
fixed value of the fifth coordinate and stretched between two D-branes.
The leading term in the large-$D$ analysis for this case turns out
to be nothing but the Arvis formula. As we will see in Sec. \ref{sec:Approximations},
when one corrects for this approximation then the leading term in
the large $D$ expansion is no longer the Arvis formula alone. Rather,
the Arvis formula gets modified by an overall factor that can be thought
of as a length-dependent string tension.

It is important to note that there should be corrections to the Arvis
formula even if the Nambu-Goto action by itself provided an exact
description of QCD flux tubes \cite{Dubovsky:2012fk}%
\footnote{I would like to thank Ofer Aharony for pointing this out to me.%
}. These expected corrections should arise from quantizing the Nambu-Goto
string in noncritical dimensions. From the point of view of the large
$D$ expansion such corrections should exist since the Arvis formula
is only the leading term in the large-$D$ expansion and there is
no \emph{a priori} reason to believe that the higher-order corrections
should vanish. Though we will not explore these higher-order corrections
in our large-$D$ analysis, but it is interesting to note that these
corrections should themselves be proportional to $D-24$ as they should
vanish for the critical dimension of $D+2=26$. Thus, surprisingly
the Arvis formula gives the exact ground-state energy of a Nambu-Goto
string in flat spacetime for $D=24$ and for $D\rightarrow\infty$.

In an alternative approach, pioneered in Refs. \cite{Aharony:2009gg,Aharony:2010cx},
one starts with a derivative expansion for the action of a fundamental
string stretched between two D-branes in five-dimensional confining
background; then one perturbatively integrates out the fluctuations
corresponding to the oscillations in the fifth dimension. These world-sheet
fluctuations are massive, and after integrating them out one obtains
an effective action for the massless transverse fluctuations. In Ref.
\cite{Aharony:2010cx} it was shown that this leads to a correction
to the Arvis potential at the order of $1/L^{4}$ , where $L$ is
the length of the fundamental string stretched between the two D-branes.
This correction comes from a boundary term induced by integrating
out the heavy modes. In another approach to boundary terms explored
in Refs. \cite{Luscher:2004ib,Billo:2012da}, one expands the boundary
action in the derivatives of the massless transverse modes. The terms
in this expansion are then constrained by the Lorentz invariance of
the underlying theory.

In our large-$D$ analysis we do not see such a boundary term at least
to the order of $1/D$, but it is worth recalling that for the physical
case of interest $D$ is equal to $2$ and therefore even $1/D^{2}$
correction can be important. What we do see is that the holographic
description of QCD flux tubes leads to additional terms to the Nambu-Goto
action that are non-zero only in the immediate vicinity of the quark
and the antiquark, and we estimate their contribution to the heavy-quark
potential. Presumably, the effects of such terms can be approximated
by introducing boundary terms in an effective string description of
a QCD flux tube.

The outline of the paper is as follows. In Sec. \ref{sec:Large D},
we consider a particular class of confining geometries and calculate
the leading term in the large-$D$ expansion under the simplifying
assumption that we discussed above. Further, in this Sec.  we recast
the large-$D$ analysis in terms of a ``master-field,'' which turns
out to be very useful for a later analysis. In Sec.  \ref{sec:1/D Correction},
within the same simplifying assumption, we calculate the $1/D$ correction
arising from the fluctuation of the string in the fifth dimension
that give rise to massive world-sheet modes. In Sec.  \ref{sec:Approximations},
we make allowance for the fact that the fundamental string descends
to the boundary smoothly over a transition region of finite length.
This leads to an effective string description of the QCD flux tube
in which the Nambu-Goto string action is supplemented by terms that
have support only in the vicinity of the quark and the antiquark.
We also estimate the contributions of these terms in the large-$D$
limit. In the final section we state our conclusions.

\section{heavy-quark Potential from Confining Geometry in the Large $D$ Limit
\label{sec:Large D}}

The gauge-invariant observable of an $\mathrm{SU(N)}$ gauge theory
that is directly related to the heavy-quark potential is the Wilson
loop, 
\begin{equation}
W\left[\Gamma\right]=\mathrm{Tr}\mathrm{\hat{\mathrm{P}}}\left(\exp i\int_{\Gamma}A\right).\label{eq:WilsonLoop-1}
\end{equation}
Here $\Gamma$ is a closed curve in four-dimensional Euclidean space,
and $A$ is the vector potential of the $\mathrm{SU(N)}$ gauge theory
\cite{Wilson:1974aa}. According to the assumed gauge/gravity duality,
the expectation value of the Wilson loop can be written as a sum over
the world-sheets of an open string whose end points terminate on the
loop $\Gamma$. The open string itself lives in a five-dimensional
curved space whose boundary is the the four-dimensional Euclidian
space where $\Gamma$ is located \cite{Witten:1998zw,Maldacena:1998im}.
Formally,

\begin{equation}
<W[\Gamma]>_{YM}\equiv\int[dA]\exp\left\{ -S[A]_{YM}\right\} =\int_{\partial X=\Gamma}[dX]\exp\left\{ -S[X]_{NG}\right\} ,\label{eq:WilsonLoop}
\end{equation}
here $S[A]_{YM}$ is the Yang-Mills action, while $X(\tau,\sigma)$
represents the world sheet of a string in a curved five-dimensional
space, and $S[X]_{NG}$ is the Nambu-Goto action. To evaluate the
Nambu-Goto action for a given surface in this space we need to know
its metric. Ideally, we would like a prescription that gives us the
geometry of the five-dimensional space once the beta function (governing
the running of the Yang-Mills coupling constant with energy) of the
dual boundary quantum field theory is specified. In the absence of
such a prescription we can look for simple modifications of $\mathrm{AdS_{5}}$
space that leads to an area law for the expectation value of the Wilson
loop. Following Ref. \cite{Polchinski:2001ju} we will consider one
such class of modification, where the metric of the curved five-dimensional
space in the coordinate system $\left\{ t,x,\Xvec,Y\right\} $ is
given by 
\begin{equation}
ds^{2}=F(Y)\left(dt^{2}+dx^{2}+d\Xvec^{2}+dY^{2}\right).\label{eq:confiningMetric-1}
\end{equation}
In this coordinate system $Y=0$ is the four-dimensional boundary%
\footnote{Strictly speaking for $AdS_{5}$ like space, $Y=0$ is a conformal
boundary \cite{Witten:1998qj}. $F(Y)$ diverges there, so in what
follows we will assume that the boundary is at $Y=\epsilon$ and that
corresponds to considering the boundary gauge-theory with a UV cutoff
of the order $1/\epsilon$ \cite{Susskind:1998dq}%
} where the gauge theory lives, $\Xvec$ denotes $D=2$ directions
transverse to the flux tube lying along the $x$ axis, and $t$ is
the Euclidian time. The area law for the expectation value of the
Wilson loop is implemented by requiring that $F(Y)$ has a minima
at some fixed value $Y=Y^{*}$. For this class of metric we will evaluate
(\ref{eq:WilsonLoop}) using a large-$D$ expansion, where $D$ is
the number of dimensions transverse to the QCD flux tube.

In the static gauge, or a physical gauge, a world-sheet of the string
connecting the quark and the antiquark is given by 
\begin{equation}
X=\left(t,x,\Xvec(t,x),Y(t,x)\right)=\left(\tau,\sigma,\Xvec\left(\vec{\sigma}\right),Y(\vec{\sigma})\right),\label{eq:staticGauge}
\end{equation}
\begin{equation}
\Xvec=\left(X^{1},X^{2},\ldots,X^{D}\right),\label{eq:Xtransverse}
\end{equation}
where we identify the world-sheet parameter $\tau$ with $t$ and
$\sigma$ with $x$. A point on the world sheet with coordinates $(\tau,\sigma)$
will be represented by $\vec{\sigma}.$ The world-sheet coordinates
range from 
\begin{equation}
-\frac{T}{2}\le t=\tau\le\frac{T}{2};\quad0\le x=\sigma\le L,\label{eq:worldsheetCoord}
\end{equation}
corresponding to a Wilson loop made up of the worldlines of a quark
at $x=0$ and an antiquark at $x=L$, and we will be interested in
the limit $T\rightarrow\infty$. Surfaces appearing in (\ref{eq:WilsonLoop})
satisfy the boundary conditions 
\begin{equation}
\Xvec\left(\tau,0\right)=0=\Xvec\left(\tau,L\right),\quad Y\left(\tau,0\right)=0=Y\left(\tau,L\right).\label{eq:WorldSheetBCond}
\end{equation}

\begin{figure}[t]
\includegraphics[scale=0.5]{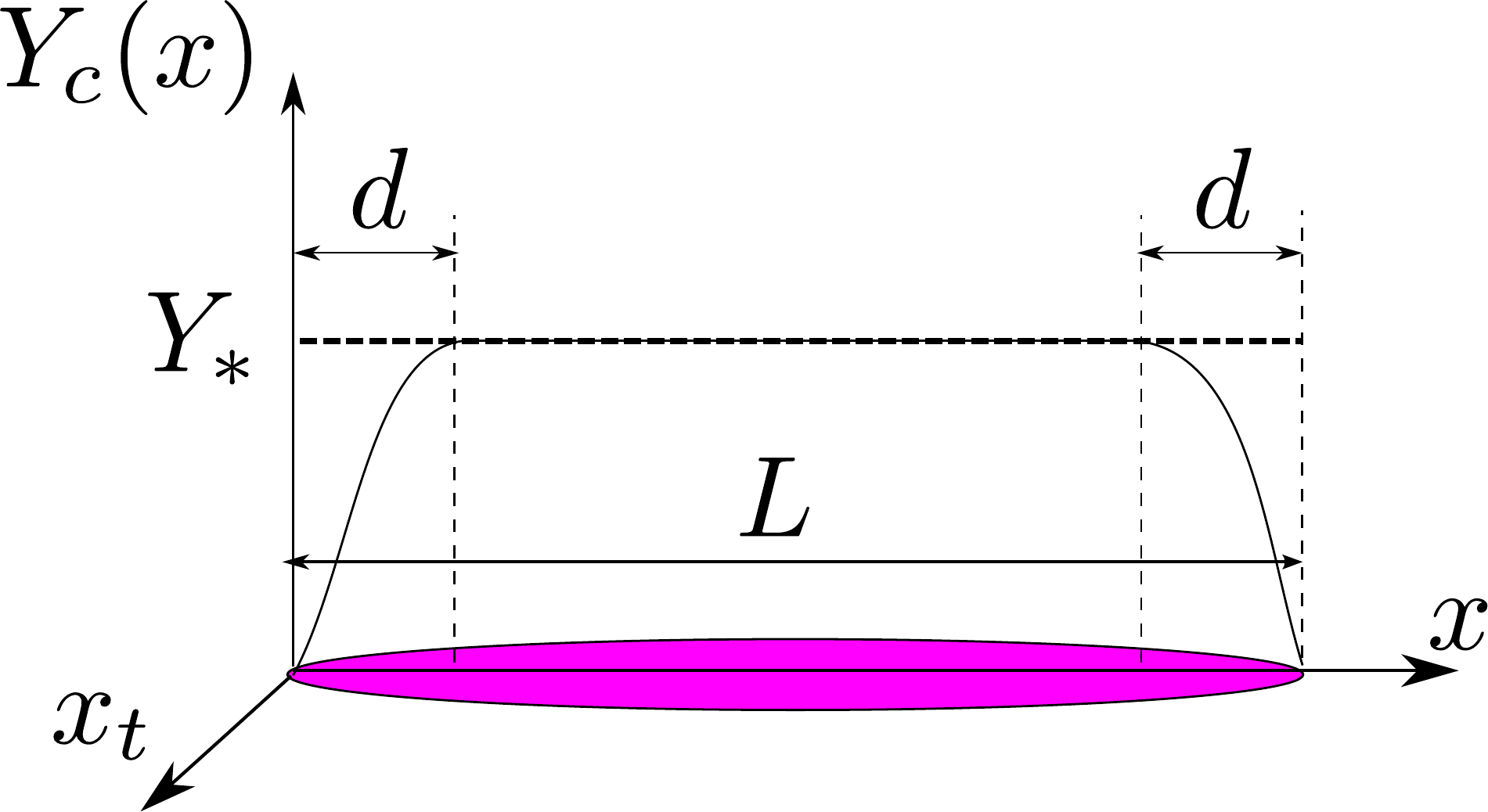}

\caption{The classical string configuration in a confining geometry \label{fig:Minimal-Surface}}
\end{figure}

The minimal surface corresponding to the metric (\ref{eq:confiningMetric-1}),
in the limit $T\rightarrow\infty$ depends only on the $x$ coordinate,
\begin{equation}
X_{c}=\left(t,x,\mathbf{0},Y_{c}\left(x\right)\right),\label{eq:minimalSurface}
\end{equation}
and is qualitatively depicted in Fig.(\ref{fig:Minimal-Surface}).
Since the warp factor $F(Y)$ has a minima at $Y=Y^{*},$ classically
the string stays at $Y=Y^{*}$ except at the end points where it dips
to connect to the quark and the antiquark that are located at $Y=0$.
The region of transition is marked $d$ in Fig.(\ref{fig:Minimal-Surface}).
For the confining geometries $d$ is expected to grow with $L$ but
with $d/L\rightarrow0$ as $L\rightarrow\infty$ \cite{Greensite:1999jw,Kinar:1999xu},
where one can define $d$, for example, as the distance for which
\begin{equation}
\frac{Y^{*}-Y_{c}(d)}{Y^{*}}=10^{-3}.\label{eq:defd}
\end{equation}

\begin{figure}
\includegraphics[scale=0.5]{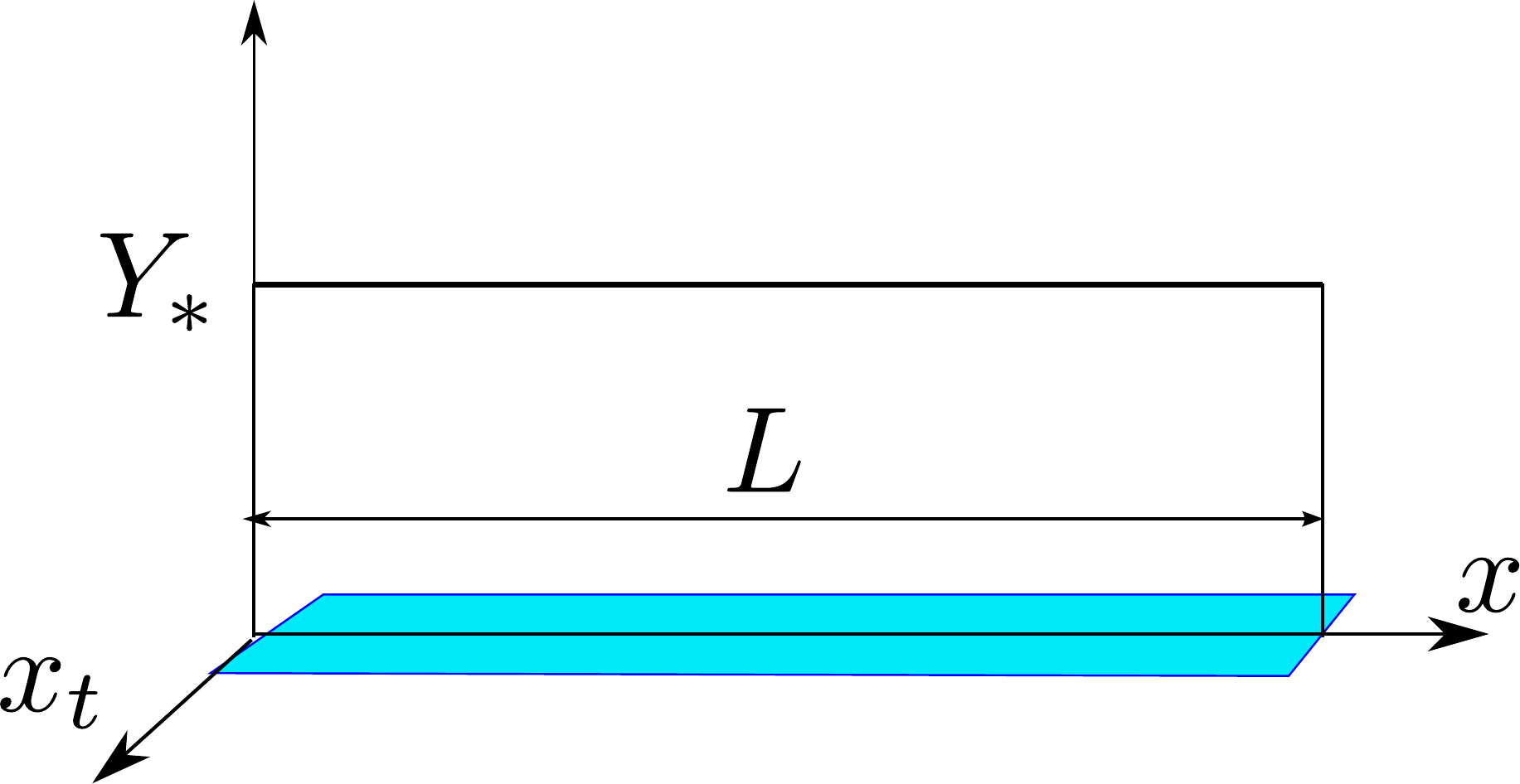}\caption{Approximation to the classical string configuration used in the large
D calculation\label{fig:approxMinSurface}}
\end{figure}

To start with, in calculating (\ref{eq:WilsonLoop}) using the large-$D$
expansion we will neglect the region $d$ and consider the fluctuation
around the surface given by 
\begin{equation}
X_{c}=\left(t,x,\mathbf{0},Y^{*}\right),\label{eq:approxMinimalSurface}
\end{equation}
as shown in Fig. \ref{fig:approxMinSurface}. In Sec. \ref{sec:Approximations}
we will return to this approximation and make an estimate of the errors
introduced by it. Within this approximation the fluctuations about
the minimal surface 
 are given by 
\begin{equation}
X=\left(t,x,\Xvec(t,x),Y^{*}+\phi(t,x)\right),\label{eq:Xfluctuations}
\end{equation}
with $\phi(x,t)$ vanishing at $x=0$ and at $x=L$. The Nambu-Goto
action for such a surface is 
\begin{equation}
S_{NG}=T_{0}\int_{-T/2}^{T/2}dt\int_{0}^{L}dx\, F(Y^{*}+\phi)\sqrt{\det\left[\gamma_{ab}\right]},\label{eq:Sng5}
\end{equation}
where $T_{0}$ is the bare string tension, and the induced world-sheet
metric $\gamma_{ab}$ (apart from the warp factor) is 
\begin{equation}
\gamma_{ab}=\delta_{ab}+\partial_{a}\Xvec\cdot\partial_{b}\Xvec+\partial_{a}\phi\partial_{b}\phi.\label{eq:inducedMetric}
\end{equation}
It will be convenient to write this metric in terms of two world-sheet
matrices, 
\begin{equation}
\boldsymbol{g}:=g_{ab}=\delta_{ab}+\partial_{a}\Xvec\cdot\partial_{b}\Xvec,\label{eq:def-g}
\end{equation}
and 
\begin{equation}
\boldsymbol{f}:=f_{ab}=\partial_{a}\phi\partial_{b}\phi.\label{eq:def-f}
\end{equation}
Then the action for the world-sheet takes the form 
\begin{equation}
S_{NG}=T_{0}\int d^{2}\sigma\, F\left(Y^{*}+\phi\right)\sqrt{\det g_{ab}}\sqrt{\det\left[\delta_{ab}+\left(g^{-1}f\right)_{ab}\right]}.\label{eq:SgabFab}
\end{equation}
One can think of $T_{0}F(Y^{*}+\phi)$ as a position-dependent string
tension. Small fluctuations in the $Y$ direction, $\phi(t,x)\ne0$
increase the string tension. These fluctuations are suppressed compared
to the fluctuations $\vec{X}\ne0,\phi=0,$ for the latter fluctuations
keep the string tension fixed at its minimum value. This suggests
the following approximate way of evaluating (\ref{eq:WilsonLoop}):
keep transverse fluctuations $\vec{X}$ to all orders but keep only
the quadratic fluctuations in $\phi$. In this approximation the Nambu-Goto
action (\ref{eq:Sng5}) takes the following form 
\begin{equation}
S_{NG5}=\frac{1}{l_{s}^{2}}\int d^{2}\sigma\sqrt{\det g_{ab}}+\frac{1}{l_{s}^{2}}\int d^{2}\sigma\sqrt{\det g_{ab}}\left(\frac{1}{2}\mathrm{Tr}\mathbf{g^{-1}f}+\frac{1}{2}M^{2}\phi^{2}\right)\label{eq:SngCorrected}
\end{equation}
where $M^{2}$ is defined via, 
\begin{equation}
F\left(Y^{*}+\phi\right)=F\left(Y^{*}\right)\left(1+\frac{1}{2}M^{2}\phi^{2}\right)+\mathcal{O}\left(\phi^{3}\right),\label{eq:defMsquare}
\end{equation}
and the fundamental length scale for the problem, $l_{s}$, is given
by 
\begin{equation}
\frac{1}{l_{s}^{2}}=T_{0}F(Y^{*}).\label{eq:defLs}
\end{equation}
In this approximation, and according to our fundamental assumption
(\ref{eq:WilsonLoop}), the expectation value of a Wilson loop is
given by

\begin{equation}
<W[\Gamma]>_{YM}=\int[d\Xvec]\exp\left\{ -\frac{1}{l_{s}^{2}}\int d^{2}\sigma\sqrt{\det g_{ab}}\right\} \mathbf{C}\left[g_{ab}\right],\label{eq:WilsonLoopPathIntegral}
\end{equation}
where 
\begin{equation}
\mathbf{C}\left[g_{ab}\right]=\int[d\phi]\exp\left\{ -\frac{1}{l_{s}^{2}}\int d^{2}\sigma\sqrt{\det g_{ab}}\left(\frac{1}{2}\mathrm{Tr}\mathbf{g^{-1}f}+\frac{1}{2}M^{2}\phi^{2}\right)\right\} .\label{eq:defC[g]}
\end{equation}

Since the heavy-quark potential, $V\left[L\right],$ is obtained from
the expectation value of a rectangular Wilson-Loop 
\begin{equation}
\lim_{T\rightarrow\infty}<W[\square_{T\times L}]>_{YM}=\mathcal{N}\exp\left\{ -TV\left[L\right]\right\} ,\label{eq:defHeavyQuarkPot}
\end{equation}
we see that $\mathbf{C}[g_{ab}]$ contains the contribution to the
heavy-quark potential due to the fluctuation of the fundamental string
in the fifth dimension. This can be heuristically interpreted as the
contribution to the heavy-quark potential due to the fluctuation of
the intrinsic thickness of the QCD flux tube.

Numerical calculations of the ground-state energy of a QCD flux tube,
in lattice gauge theory matches very well with the ground-state energy
of the Nambu-Goto string in four Euclidean dimensions as given by
the Arvis formula. This suggests that the fluctuations of the string
in the fifth dimension should only provide a small correction to the
ground-state energy given by the Nambu-Goto string in the four flat
dimensions. Therefore, setting $C[\boldsymbol{g}]=\mathrm{constant}$
in (\ref{eq:WilsonLoopPathIntegral}) which corresponds to describing
the QCD flux tube by a Nambu-Goto string in four Euclidean dimensions
and ignoring the fluctuations in the intrinsic thickness--is a good
approximation.

The above observations can be made more precise by evaluating (\ref{eq:WilsonLoopPathIntegral})
in the large-$D$ expansion, where $D$ is the number of transverse
dimensions of the $D+2$-dimensional boundary. As we will see, in
the large-$D$ expansion of (\ref{eq:WilsonLoopPathIntegral}) the
smallness of the correction due to $C[\boldsymbol{g}]$ will be the
consequence of the fact that there is only one massive $\phi$ field
while there are $D$ massless transverse fields, $X_{i}$.

The large-$D$ expansion will be implemented in the standard manner
\cite{Alvarez:1981kc}. We start with the path integral 
\begin{equation}
Z_{5}=\int[d\Xvec]\exp\left\{ -\frac{1}{l_{s}^{2}}\int d^{2}\sigma\sqrt{\det g_{ab}}\right\} \mathbf{C}\left[g_{ab}\right],\label{eq:defZ5}
\end{equation}
and introduce $\boldsymbol{g}$ as an independent degree of freedom
in the path integral but with the constraint (\ref{eq:def-g}). Next
we implement the constraint using the Lagrangian multiplier field
$N$. This converts the integration over $\Xvec$ fields into Gaussian
integrals and one obtains 
\begin{equation}
Z_{5}=\mathcal{N}_{1}\int\left[d\boldsymbol{g}\right]\left[d\boldsymbol{N}\right]\exp\left\{ -S_{eff}\left[\boldsymbol{g},\boldsymbol{N}\right]\right\} \mathbf{C}\boldsymbol{[g]},\label{eq:Z5AuxFields}
\end{equation}
where 
\begin{equation}
S_{eff}[\mathbf{g},\mathbf{N}]=\frac{1}{l_{s}^{2}}\int d^{2}\sigma\left\{ \sqrt{\det\mathbf{g}}+\frac{1}{2}N^{ab}(\delta_{ab}-g_{ab})\right\} +\frac{D}{2}\mathrm{Tr}\left[\log\left(-\partial_{a}N^{ab}\partial_{b}\right)\right],\label{eq:S[g,N]}
\end{equation}
and $\mathcal{N}_{1}$ is a normalization constant. Further, we can
integrate over the $\phi$ field in (\ref{eq:defC[g]}) and write
\begin{equation}
\mathbf{C}[\boldsymbol{g}]=\mathcal{N}_{2}\exp\left\{ -\frac{1}{2}\mathrm{Tr}\left[\log\left(\hat{\mathbf{A}}\right)\right]\right\} ,\label{eq:Seff2(g)}
\end{equation}
where 
\begin{equation}
\hat{\mathbf{A}}\left(\boldsymbol{g}\right)=-\partial_{a}\sqrt{\det\boldsymbol{g}}g_{ab}^{-1}\partial_{b}+\sqrt{\det\boldsymbol{g}}M^{2},\label{eq:defAhat}
\end{equation}
and $\mathcal{N}_{2}$ is another normalization constant.

We would like to evaluate (\ref{eq:defZ5}) in the limit $\mathrm{D}\rightarrow\infty$
while keeping $Dl_{s}^{2}$ constant. In other words we are interested
in the limit 
\begin{equation}
\mathrm{D}\rightarrow\infty;\quad\Lbarsquare=\mathrm{constant,}\label{eq:largeDlimit}
\end{equation}
where $\Lbarsquare$ is defined as 
\begin{equation}
\Lbarsquare=Dl_{s}^{2}.\label{eq:defLsbar}
\end{equation}
The path integral (\ref{eq:defZ5}) can now be written in the following
form 
\begin{equation}
Z_{5}=\mathcal{N}\int\left[d\boldsymbol{g}\right]\left[d\boldsymbol{N}\right]\exp\left\{ -D\, S\left[\boldsymbol{g},\boldsymbol{N}\right]\right\} ,\label{eq:Z5LargeD}
\end{equation}
where $S\left[\boldsymbol{g},\boldsymbol{N}\right]$ has two parts,
\begin{equation}
S\left[\boldsymbol{g},\boldsymbol{N}\right]=\left(\, S_{1}\left[\boldsymbol{g},\boldsymbol{N}\right]+\frac{1}{D}S_{2}\left[\boldsymbol{g}\right]\right),\label{eq:defS[g,N]}
\end{equation}
with 
\begin{equation}
S_{1}\left[\boldsymbol{g},\boldsymbol{N}\right]=\frac{1}{\Lbarsquare}\int d^{2}\sigma\left\{ \sqrt{\det\mathbf{g}}+\frac{1}{2}N^{ab}(\delta_{ab}-g_{ab})\right\} +\frac{1}{2}\mathrm{Tr}\left[\log\left(-\partial_{a}N^{ab}\partial_{b}\right)\right],\label{eq:defS1}
\end{equation}
and 
\begin{equation}
S_{2}\left[\boldsymbol{g}\right]=\frac{1}{2}\mathrm{Tr}\left[\log\left(\hat{\mathbf{A}}\left(\boldsymbol{g}\right)\right)\right].\label{eq:defS2}
\end{equation}
Therefore in the large-$D$ limit, the contribution of the fluctuations
of the $\phi$ field is suppressed by a factor of $1/D$ as compared
to the contributions of the fluctuations of the transverse coordinates
$X_{i}$. From the point of view of the large-$D$ expansion, the
success of the four-dimensional Nambu-Goto string in reproducing the
heavy-quark potential is due to the fact that the corrections to it
are suppressed by a factor of $1/D.$

The leading term in the large-$D$ expansion of (\ref{eq:Z5LargeD})
is then given by 
\begin{equation}
Z_{5}=\mathcal{N}_{5}\exp\left\{ -D\, S_{1}\left[\bar{\boldsymbol{g}},\bar{\boldsymbol{N}}\right]\right\} ,\label{eq:myLargeDapprox}
\end{equation}
where the $\bar{\boldsymbol{g}}$ and $\bar{\boldsymbol{N}}$ are
the fields that minimize the action $S_{1}$ alone, 
\begin{equation}
\frac{\delta S_{1}}{\delta\boldsymbol{g}}=0,\quad\frac{\delta S_{1}}{\delta\boldsymbol{N}}=0.\label{eq:masterFields}
\end{equation}
Minimizing the action $S_{1}$ alone is nothing but the large-$D$
limit of the Nambu-Goto string in a flat Euclidean space \cite{Alvarez:1981kc}
and the required solutions are: 
\begin{eqnarray}
\bar{\boldsymbol{g}} & =\left[\begin{array}{cc}
g_{1} & 0\\
0 & g_{2}
\end{array}\right]= & \begin{bmatrix}\frac{1-\lambda}{1-2\lambda} & 0\\
0 & 1-\lambda
\end{bmatrix},\label{eq:gbar}\\
\bar{\boldsymbol{N}} & =\left[\begin{array}{cc}
N_{1} & 0\\
0 & N_{2}
\end{array}\right]= & \begin{bmatrix}\sqrt{1-2\lambda} & 0\\
0 & \frac{1}{\sqrt{1-2\lambda}}
\end{bmatrix},\label{eq:Nbar}
\end{eqnarray}
where $\lambda$ depends only on the distance between the quark and
the antiquark, $L$, and is given by 
\begin{equation}
\lambda\left(L\right)=\frac{\pi}{24}\frac{\Lbarsquare}{L^{2}}=\frac{\pi D}{24}\frac{l_{s}^{2}}{L^{2}}.\label{eq:lambda}
\end{equation}
A fact that will be important for us latter--when we calculate the
$1/D$ corrections--is that the metric $\bar{\boldsymbol{g}}$$ $
has discontinuities at the boundaries--$\sigma=x=0$ and $ $ $\sigma=x=L$,
which are the worldlines of the quark and the antiquark (see Appendix
of Ref. \cite{Alvarez:1981kc}.) The metric at the boundary is given
by 
\begin{equation}
\bar{\boldsymbol{g}}_{b}=\left[\begin{array}{cc}
g_{b1} & 0\\
0 & g_{b2}
\end{array}\right]=\left[\begin{array}{cc}
1 & 0\\
0 & 1-2\lambda
\end{array}\right],\label{eq:gbarBoundary}
\end{equation}
but, interestingly, $N$ is continuous, 
\begin{equation}
\bar{\boldsymbol{N}}=\left[\begin{array}{cc}
\sqrt{\frac{g_{b2}}{g_{b1}}} & 0\\
0 & \sqrt{\frac{g_{b1}}{g_{b2}}}
\end{array}\right]=\left[\begin{array}{cc}
\sqrt{1-2\lambda} & 0\\
0 & \frac{1}{\sqrt{1-2\lambda}}
\end{array}\right].\label{eq:Nbargbar}
\end{equation}

$ $Using the above results, one readily obtains the ground-state
energy of a Nambu-Goto string in the large-$D$ limit \cite{Alvarez:1981kc},
which turns out to be the Arvis formula for the ground-state energy
of a Nambu-Goto string \cite{Arvis:1983fp}, 
\begin{equation}
DS_{1}\left[\bar{\boldsymbol{g}},\bar{\boldsymbol{N}}\right]=TV_{A}\left[L\right],\label{eq:S1bar}
\end{equation}
\begin{equation}
V_{A}\left[L\right]=\frac{L}{l_{s}^{2}}\sqrt{1-2\lambda\left(L\right)}.\label{eq:arvisPot}
\end{equation}

The large-$D$ expansion bears a close resemblance to the large-$N$
expansion of multi-component quantum field theories (see, e.g., Ref.
\cite{coleman:aspects} for a review). It is also similar in spirit
to the mean-field approximation (for a mean-field analysis of the
Nambu-Goto string see \cite{Makeenko:2012fk}). Continuing with the
analogy with the large-$N$ expansion further, one can ask if there
is a single surface that saturates the partition function of the Nambu-Goto
string in the $D\rightarrow\infty$ limit. Such a surface would be
the analog of the master field of the large-$N$ gauge theories. In
fact there is such a fictitious mathematical surface--whose metric
is given by the boundary metric $\bar{\mathbf{g}}_{b}$ (\ref{eq:gbarBoundary}),

\begin{eqnarray}
Z_{NG} & = & \int[d\Xvec]\exp\left\{ -\frac{1}{l_{s}^{2}}\int d^{2}\sigma\sqrt{\det g_{ab}}\right\} =\mathcal{N}_{NG}\exp\left\{ -\frac{1}{l_{s}^{2}}\int_{-T/2}^{T/2}dt\int_{0}^{L}dx\sqrt{\det\bar{\mathbf{g}}_{b}}\right\} ,\nonumber \\
 & = & \mathcal{N}_{NG}\exp\left\{ -\frac{TL}{l_{s}^{2}}\sqrt{1-2\lambda(L)}\right\} .\nonumber \\
\label{eq:masterField}
\end{eqnarray}
This observation will be of great use to us in evaluating the $1/D$
corrections in the next Sec. . The metric $ $$\bar{\mathbf{g}}_{b}$
should not be thought of as an induced metric, but as an independent
metric defining the ``master surface''. It is worth noting that
$ $$g_{b1}$ coincides with the induced metric on the quark and antiquark
worldline (apart from the overall warp factor that is absorbed in
our definition of $l_{s}$).

\section{The $1/D$ Corrections from Heavy Modes\label{sec:1/D Correction}}

We will evaluate the $1/D$ corrections arising from the oscillation
of the string along the fifth dimension using the observation made
in the last section , namely that in the large-$D$ limit the partition
function of the Nambu-Goto string is given by a single ``master surface''
with metric given by (\ref{eq:gbarBoundary}). The partition function
of the fundamental string in five dimensions, (\ref{eq:defZ5}), can
be written as 
\begin{equation}
Z_{5}=Z_{NG}<\mathbf{C}[\boldsymbol{g}]>_{NG}=\mathcal{N}_{2}Z_{NG}<\exp\left\{ -\frac{1}{2}\mathrm{Tr}\left[\log\left(\hat{\mathbf{A}}[\mathbf{g]}\right)\right]\right\} >_{NG}.\label{eq:1/DCorrUsingNG}
\end{equation}
In the large-$D$ limit, this average in turn can be evaluated as
\begin{equation}
Z_{5}=\mathcal{N}_{5}\exp\left\{ -\frac{TL}{l_{s}^{2}}\sqrt{1-2\lambda\left(L\right)}\right\} \exp\left\{ -\frac{1}{2}\mathrm{Tr}\left[\log\left(\hat{\mathbf{A}}[\bar{\mathbf{g}}_{b}]\right)\right]\right\} ,\label{eq:1/DcorrFromMasterField}
\end{equation}
where, using (\ref{eq:defAhat}) and (\ref{eq:gbarBoundary}), 
\begin{equation}
\hat{\mathbf{A}}\left[\bar{\boldsymbol{g}}_{b}\right]=-\sqrt{\frac{g_{b2}}{g_{b1}}}\partial_{1}^{2}-\sqrt{\frac{g_{b1}}{g_{b2}}}\partial_{2}^{2}+\sqrt{\det\bar{\boldsymbol{g}}_{b}}M^{2}\label{eq:defAgbar}
\end{equation}
The trace can be calculated using zeta-function regularization in
a standard manner (see, e.g., Ref. \cite{Nesterenko:1997ku}), 
\begin{equation}
-\frac{1}{2}\mathrm{Tr}\left[\log\left(\hat{\mathbf{A}}[\bar{\mathbf{g}}_{b}]\right)\right]=-TV_{M}\left(L\right),\label{eq:defVc1}
\end{equation}
where 
\begin{equation}
V_{M}(L)=\left(-\frac{1}{4}g_{b1}^{1/4}M+\frac{1}{4\pi}\sqrt{\frac{g_{b1}}{g_{b2}}}\int_{0}^{\infty}d\omega\log\left(1-\mathrm{e}^{-2L\sqrt{\omega^{2}+\omega_{0}^{2}}}\right)\right),\label{eq:Vc1a}
\end{equation}
and 
\begin{equation}
\omega_{0}=g_{b2}^{1/2}M.\label{eq:defw0}
\end{equation}
Using (\ref{eq:gbarBoundary}) we get 
\begin{equation}
V_{M}\left(L\right)=-\frac{1}{4}M+V_{h}(L),\label{eq:v1cb}
\end{equation}
where 
\begin{equation}
V_{h}(L)=\frac{1}{4\pi}\frac{1}{\sqrt{1-2\lambda}}\int_{0}^{\infty}d\omega\log\left(1-\mathrm{e}^{-2L\sqrt{\omega^{2}+\left(1-2\lambda\right)M^{2}}}\right).\label{eq:defVh}
\end{equation}
For $LM>>1$, the integral leads to an exponential decaying term,
while for $LM<<1$ one would obtain an additional $1/L$ correction
to the heavy-quark potential. Such a term is of course precluded by
both theoretical arguments \cite{Luscher:2004ib} and numerical results.
In any case, for $LM<<1$ our large-$D$ analysis fails, for then
the approximation of replacing $F[Y_{c}]$ by $F[Y^{*}]$ in the Nambu-Goto
action is no longer valid, equivalently, the approximation $d=0$
in Fig. \ref{fig:Minimal-Surface} is no longer valid. Therefore,
the only region where the above correction is of significance is when
$LM\sim1$, which we can evaluate numerically. 
\begin{figure}
\includegraphics[scale=1.45]{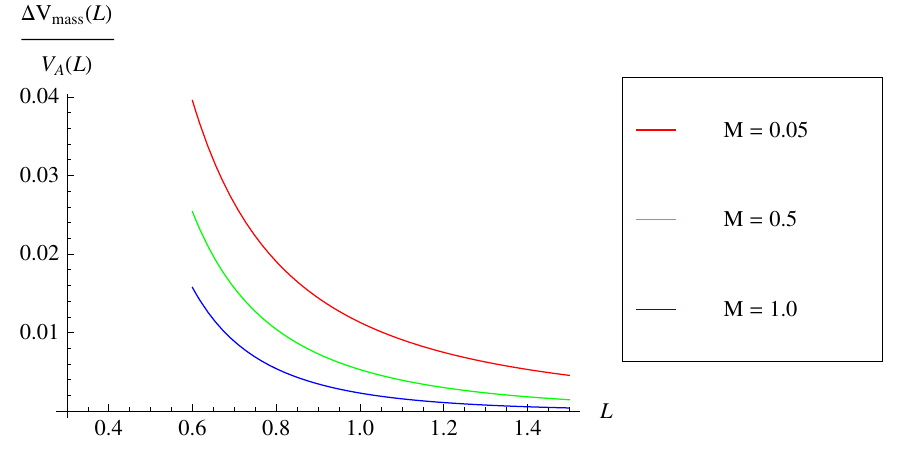}

\caption{Effect of a massive mode on the heavy-quark potential, where $\Delta V_{\textrm{mass}}\equiv|V_{A}-V_{h}|$\label{fig:massModeContrb}}
\end{figure}

In Fig. \ref{fig:massModeContrb} we show the contribution of the
difference between $V_{A}(L)$ and $V_{h}(L)$ as compared to the
Arvis potential $V_{A}(L)$ for representative values of $M$ in the
units of inverse fermi using $l_{s}=0.4$ fermi. Therefore, the main
qualitative contribution of the heavy modes is to modify the Arvis
potential by the addition of a constant term, $-\frac{1}{4}M$.

\section{Massless Modes Near the Boundary \label{sec:Approximations}}

Our large-$D$ calculation has been done under the approximation in
which the minimal surface of the fundamental string is taken to be
(\ref{eq:approxMinimalSurface}), $X_{c}=(t,x,\vec{\mathbf{0}},Y^{*}),$
that amounts to taking $d=0$ in Fig.(\ref{fig:Minimal-Surface}).
As a result of this approximation, in evaluating (\ref{eq:WilsonLoop})
the fluctuations in the string world sheets are incorrectly weighted
near the vicinity of the quark and the antiquark. If we relax the
approximation $d=0$ {[}see Fig. \ref{fig:Minimal-Surface} and Fig.
\ref{fig:approxMinSurface}{]}, then the minimal surface is given
by (\ref{eq:minimalSurface}), $X_{c}=(t,x,\vec{\mathbf{0}},Y_{c}\left(x\right)).$
We would like to consider the fluctuations around the exact minimal
surface; in particular we will consider the fluctuations around it
of the form 
\begin{equation}
X=\left(t,x,\Xvec\left(t,x\right),Y_{c}(x)\right),\label{eq:fluctOfMinSurface}
\end{equation}
that is we will be ignoring the fluctuations of the string along the
fifth dimension that are massive, but will include the massless fluctuation
around the exact minimal surface. In the static gauge, the action
for the above fluctuations is given by 
\begin{equation}
S[\Xvec;F[Y_{c}(x)]]=T_{0}\int dtdxF\left[Y_{c}\left(x\right)\right]\sqrt{\det g_{ab}}\cdot\left(1+g_{22}^{-1}\left(\partial_{x}Y_{c}\right)^{2}\right)^{1/2},\label{eq:NGincludingBoundary}
\end{equation}
where the world-sheet metric $g_{ab}$ is given by (\ref{eq:def-g})
and $T_{0}$ is the bare fundamental string tension. To delineate
the effects of the boundary one can rewrite the above action as 
\begin{align}
S\left[\Xvec;F[Y_{c}(x)]\right]= & T_{0}F(Y^{*})\int_{-T/2}^{T/2}dt\int_{0}^{L}dx\sqrt{\det g_{ab}}\nonumber \\
+ & T_{0}\int_{-T/2}^{T/2}dt\int_{x\in\mathbf{B}}dx\sqrt{\det g_{ab}}\left\{ F\left[Y_{c}\left(x\right)\right]\left(1+g_{22}^{-1}\left(\partial_{x}Y_{c}\right)^{2}\right)^{1/2}-F(Y^{*})\right\} ,\label{eq:separatingBoundaryTerms}
\end{align}
where we have used the fact that in the interval $\left(d,\, L-d\right)$,
to a very good approximation $Y_{c}(x)=Y^{*}$. Also we have represented
the boundary regions, $(0,\, d)$ and $\left(L-d,\, L\right)$, by
$\mathbf{B}$.

Since the warp factor $F\left[Y_{c}(x)\right]$ differs from $F[Y^{*}]$
only near the vicinity of the quark and the antiquark, we rewrite
it as 
\begin{equation}
F[Y_{c}(x)]=F(Y^{*})\left(1+f_{c}(x)\right),\quad f_{c}(x)=\frac{F_{c}[Y_{c}(x)]}{F[Y^{*}]}-1,\label{eq:deff_c(x)}
\end{equation}
and the action for the massless fluctuations becomes 
\begin{multline}
S\left[\Xvec;F[Y_{c}(x)]\right]=\frac{1}{l_{s}^{2}}\int_{-T/2}^{T/2}dt\int_{0}^{L}dx\sqrt{\det g_{ab}}\\
+\frac{1}{l_{s}^{2}}\int_{-T/2}^{T/2}dt\int_{x\in\mathbf{B}}dx\sqrt{\det g}\left\{ \left(1+g_{22}^{-1}\left(\partial_{x}Y_{c}\right)^{2}\right)^{1/2}-1\right\} \\
+\frac{1}{l_{s}^{2}}\int_{-T/2}^{T/2}dt\int_{x\in\mathbf{B}}dx\sqrt{\det g_{ab}}f_{c}(x)\left(1+g_{22}^{-1}\left(\partial_{x}Y_{c}\right)^{2}\right)^{1/2},\label{eq:boundaryTerms}
\end{multline}
where we have used $T_{0}F(Y*)=1/l_{s}^{2}$.

The first term in the above equation is nothing but the Nambu-Goto
action for a QCD flux tube. The rest of the terms, which originate
from the holographic description of QCD flux tube, are corrections
to the Nambu-Goto action. These terms have support only in the vicinity
of the quark and the antiquark, a region that we have denoted by $\mathbf{B}$.
Let us first consider the last term. Near the boundary, $Y_{c}(x)\rightarrow0$,
the string experiences $\mathrm{AdS_{5}}$ like geometry (that ensures
that at very short distances the heavy-quark potential is coulomb
like); as a result the warp factor $f_{c}[Y_{c}(x)]$ behaves like
\begin{equation}
\lim_{Y_{c}[x]\rightarrow0}f_{c}[Y_{c}(x)]\approx\frac{1}{F(Y^{*})}\frac{R^{2}}{Y_{c}(x)^{2}}\approx\frac{1}{F(Y^{*})}\frac{R^{2}}{\left(mx\right)^{2}},\label{eq:fcNearBoundary}
\end{equation}
where we have focused on the $x=0$ end of the flux tube. $R$ is
the radius of curvature of the $\mathrm{AdS_{5}}$ space near the
boundary and $m=\partial_{x}Y_{c}(0)$. As a result the last term
in (\ref{eq:boundaryTerms}) is divergent. To understand the nature
of this divergence, let us put the boundary at $Y=m\epsilon$ and
evaluate this term for the classical configuration (\ref{eq:minimalSurface})
\begin{equation}
\lim_{\epsilon\rightarrow0}T_{0}\int dt\int_{\epsilon}^{d}dx\frac{R^{2}}{x^{2}}\left(1+\left(\partial_{x}Y_{c}\right)^{2}\right)^{1/2}\approx T_{0}\int dt\frac{R^{2}\mathbf{C}(0)}{\epsilon},\label{eq:boundaryDIvergence}
\end{equation}
where $\mathbf{C}(x)=\left(1+\left(\partial_{x}Y_{c}\right)^{2}\right){}^{1/2}$
is regular at $x=0$. Placing the boundary of the five-dimensional
space at $Y=m\epsilon$ instead of at $Y=0$ is equivalent to introducing
an ultraviolet cutoff in the boundary theory and the divergence of
the last term corresponds to the linearly divergent contribution to
the self-energy of the quark \cite{Susskind:1998dq}, 
\begin{equation}
\delta M=T_{0}\frac{R^{2}\mathbf{C}(0)}{\epsilon}.\label{eq:divergentQuarkMass}
\end{equation}
 From the flux tube point of view the divergence arises because near
the quark and the antiquark the intrinsic thickness is going to zero
and correspondingly the effective string tension diverges \cite{Vyas:2010uq}.
Thus, the main effect of the last term in (\ref{eq:boundaryTerms})
is just to renormalize the mass of the quark and the antiquark on
which the string terminates, and we will therefore ignore it in the
discussion below.

\begin{figure}
\includegraphics[scale=1.25]{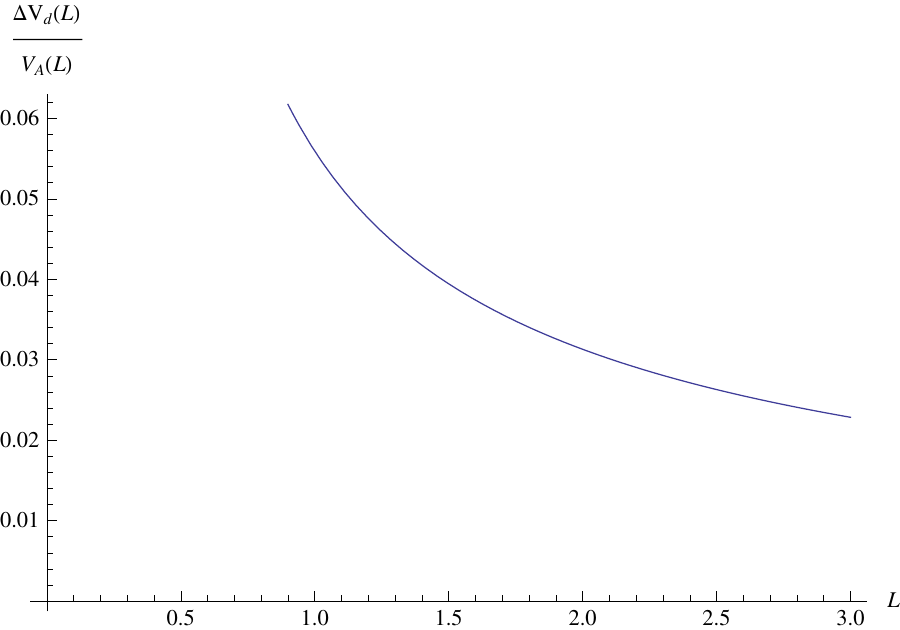}

\caption{The effect of massless modes near the boundaries on the heavy-quark
potential, where $\Delta V_{d}\equiv|V_{A}-V_{d}|$\label{fig:TheBoundaryEffect}}
\end{figure}

The second term in (\ref{eq:boundaryTerms}) is finite and represents
a correction to the Nambu-Goto string in four dimensions. We can estimate
its effect on the heavy-quark potential by treating it as a perturbation
to the Nambu-Goto action, 
\begin{eqnarray}
Z_{d} & \equiv & \int\left[d\Xvec\right]\exp\left\{ -\frac{1}{l_{s}^{2}}\int_{-T/2}^{T/2}dt\int_{0}^{L}dx\sqrt{\det g_{ab}}-\frac{1}{l_{s}^{2}}\int_{-T/2}^{T/2}dt\int_{x\in\mathbf{B}}dx\mathbf{b}(t,x)\right\} \nonumber \\
 & \approxeq & Z_{NG}\left\langle \exp\left(-\frac{1}{l_{s}^{2}}\int_{-T/2}^{T/2}dt\int_{x\in\mathbf{B}}dx\mathbf{b}(t,x)\right)\right\rangle _{NG},\label{eq:Zd}
\end{eqnarray}
where we have defined, 
\begin{equation}
\mathbf{b}(t,x)\equiv\sqrt{\det g}\left\{ \left(1+g_{22}^{-1}\left(\partial_{x}Y_{c}\right)^{2}\right)^{1/2}-1\right\} .\label{eq:defB}
\end{equation}
We can evaluate (\ref{eq:Zd}) in the large-$D$ limit using (\ref{eq:masterField})
and by approximating $\partial_{x}Y_{c}(x)\approx\partial_{x}Y_{c}(0),$
to obtain the heavy-quark potential as 
\begin{equation}
V_{d}[L]=\left(1+h\left(L\right)\right)V_{A}[L],\label{eq:boundCorrPot}
\end{equation}
where $V_{A}[L]$ is the Arvis potential (\ref{eq:arvisPot}) and
\begin{equation}
h(L)=\frac{2d}{L}\left\{ \left(1+\frac{\left(\partial_{x}Y_{c}(0)\right)^{2}}{1-2\lambda(L)}\right)^{1/2}-1\right\} .\label{eq:defh}
\end{equation}
In \cite{Kinar:1999xu} it was argued that for confining geometries
$d$ is a function of $L$ and grows with $L$ either as a power law
or logarithmically, but with $d/L\rightarrow0$ as $L\rightarrow\infty$.
With this in mind, and to get a qualitative feeling for the corrections
to the Arvis potential, we parametrize 
\begin{equation}
2d=d_{0}\left(\frac{L}{l_{s}}\right)^{\beta},\label{eq:dAsFuncOfL}
\end{equation}
where $0<\beta<1$ and take $d_{0}$ to be the inverse glueball mass,
which is a measure of the correlation length of the gauge field. To
plot these corrections, we will arbitrarily take the value of $\beta=1/2$,
and approximate the classical string configuration near the boundary
by a straight line, $(\partial_{x}Y_{c}(0))^{2}\sim1,$ and take $m_{\textrm{glueball}}\approx2GeV$
and $l_{s}=0.4$ fermi. With these parameters the relative deviation
from the Arvis potential is plotted in Fig. (\ref{fig:TheBoundaryEffect})
and is, at least for our choice of parameters, significantly more
than the corresponding correction due to heavy modes, Fig. (\ref{fig:massModeContrb}).

\begin{figure}
\includegraphics[scale=0.45]{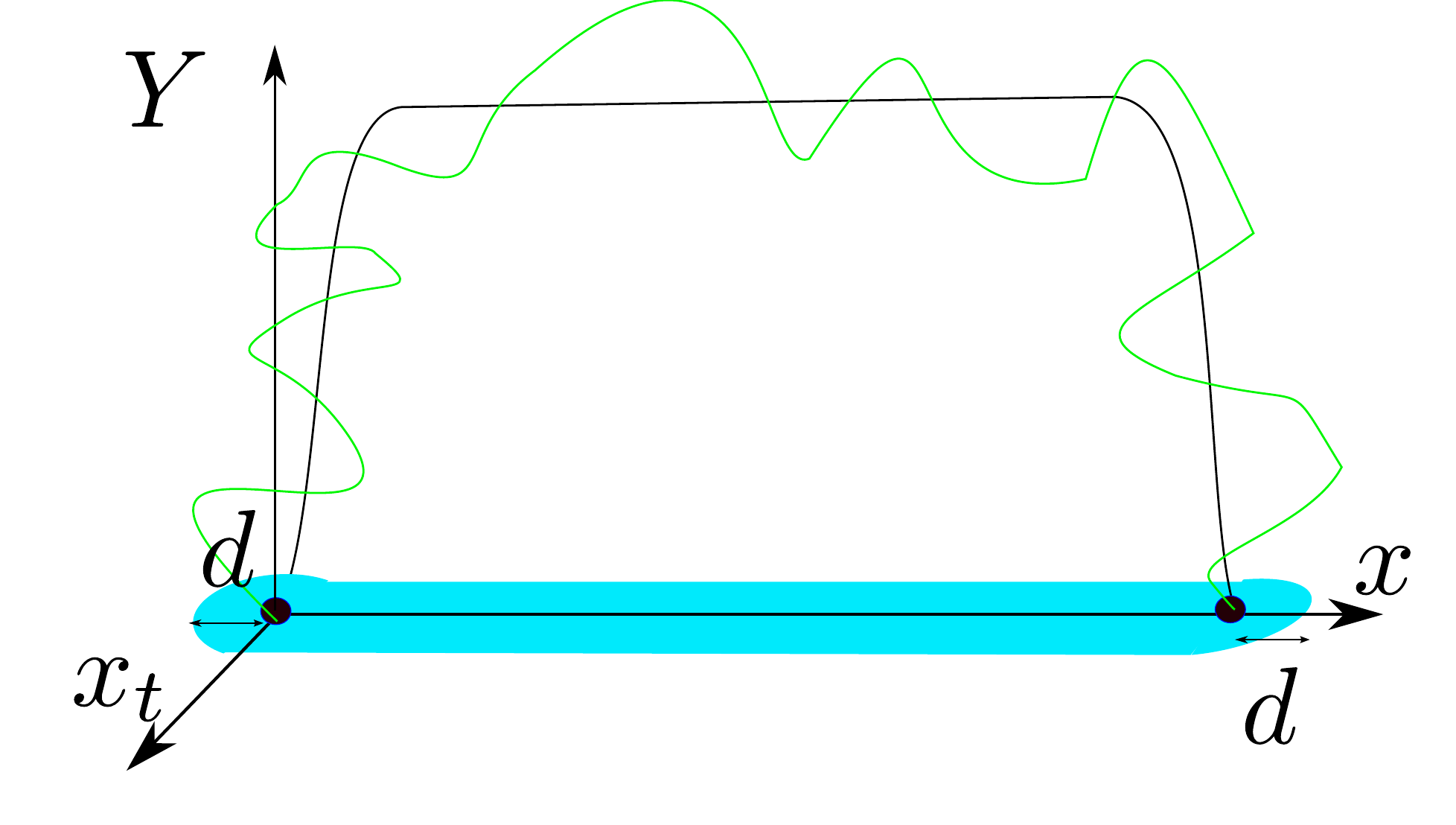}

\caption{Fluctuation of string in the extra dimension about the classical solution\label{fig:FluctLeadingToCap}}
\end{figure}

Finally, let us take note of the fluctuations like the one depicted
in Fig. \ref{fig:FluctLeadingToCap}. These are the massive fluctuations
of the string in the extra dimension about the classical solution
$Y_{c}(x)$. These fluctuations differ from the fluctuations that
we have considered in Sec.  {[}\ref{sec:1/D Correction}{]} only near
the boundary. It is worth noting that these fluctuations are most
conveniently described using a normal gauge (see, e.g., \cite{Kinar:1999xu})
rather than in the $(t,x)$ gauge that we have used, for when $x$
is close to the boundary even a small fluctuation $\phi(t,x)$ in
the $Y$ coordinate can become a multiple-valued function of $x$.
In the holographic description of QCD flux tube these fluctuations
are suggestive of the \emph{longitudinal} fluctuations causing the
flux tube to extend beyond the quark and the anti quark. In a future
work we hope to probe their effects in confining geometries.

\section{Summary and Conclusions\label{sec:Summary-and-Conclusions}}

The key features of the heavy-quark potential as suggested by the
gauge/gravity duality in the large-$D$ limit can be summarized as
\begin{equation}
V[L]=\left(1+h(L)\right)\frac{L}{l_{s}^{2}}\sqrt{\left(1-\frac{2\pi}{12}\frac{l_{s}^{2}}{L^{2}}\right)}+\left(-\frac{1}{4}M+V_{h}[L]\right),\label{eq:sumCorrection}
\end{equation}
where $h(L)$ is given by (\ref{eq:defh}) and arises from the behavior
of the massless modes in the vicinity of the quark and the antiquark.
The contribution from the heavy modes is contained in $V_{h}[L]$,
which is given by (\ref{eq:defVh}) and its relative effect on the
heavy-quark potential is shown in Fig. \ref{fig:massModeContrb}.
The leading contribution to the heavy-quark potential in the large-$D$
expansion is contained in the first term, while the last parentheses
contains the $1/D$ corrections arising from the oscillations of the
fundamental string in the fifth dimension. As we have noted in the
Introduction, there are possible $1/D$ and higher corrections arising
from the massless modes in noncritical dimensions, which we have ignored
in our analysis. To understand the manner in which they vanish for
the critical dimension is an important and interesting problem to
which we hope to come to in the future.

There are two key qualitative features of $V[L]$. Firstly, there
is $ $the correction to the Arvis potential by the factor $h(L)$.
Even though in our investigation we cannot fix the precise form of
$h(L)$, we can clearly see the reason for its existence. It arises
from the simultaneous requirement that the fundamental string wants
to lie at a fixed value $Y=Y^{*}$, to minimize its energy, and the
requirement that the fundamental string has to terminate on the boundary,
$Y=0$, where the quark and the antiquark are placed. Thus, the origin
of $h(L)$ is in the behavior of the flux tube in the vicinity of
the quark and the antiquark. Though the present numerical results
are consistent with $h(L)=0,$ but our analysis strongly suggests
that there should be corrections to the Arvis potential, especially
around $L\sim1\,\mathrm{fermi}$ and they should become visible in
future calculations of the heavy-quark potential on the lattice. The
other interesting feature of (\ref{eq:sumCorrection}) is the existence
of the constant term in the potential, $-M/4$. At first sight the
existence of the constant term in the potential is of no consequence,
but--as has been emphasized in Refs. \cite{Hidaka:2009xh,Kol:2010fq}--once
the renormalization prescription for the expectation value of the
Wilson loop is appropriately fixed, the constant arises only from
the corrections to the Nambu-Goto string %
\footnote{I would like to thank R. D. Pisarski for pointing this out to me.%
}. It is worth recalling that there is no constant term in the ground-state
energy of a Nambu-Goto string as given by the Arvis formula.

The main motivation behind calculating the heavy-quark potential using
gauge/gravity duality is to compare its predication with that of lattice
QCD at all scales; that is, from the scale where the potential is
Coulombic and the coupling constant is weak, to the strong-coupling
scale where the QCD flux tubes are formed. The hope is that in this
manner we can delineate the geometry of the five-dimensional curved
space that incorporates both asymptotic freedom and confinement. Having
found such a geometry one can hope to calculate more interesting quantities,
like the hadron masses and their parton distributions.

\section*{Acknowledgments}

I would like to thank Ofer Aharony for his very useful comments on
a preliminary version of this paper. I have also greatly benefited
from my discussions with the participants of the ECT{*} workshop on
\textquotedblleft{}Confining flux tubes and strings\textquotedblright{}
in Trento, 2010, and would like to thank them all. This work was started
while I was visiting the Institute of Theoretical Physics, at the
University of Regensburg. I am grateful to the members of the Institute
for their hospitality, particularly to Gunnar Bali for his invitation
to visit and DAAD for their financial support. Part of this work was
done subsequently while I was visiting the Raman Research Institute,
Bangalore and I would like to thank the members of the theoretical
physics group at RRI for their hospitality and Madhavan Varadarajan
for his kind invitation. I would also like to thank Janaki Abraham
for her help in proof reading this paper.

\providecommand{\href}[2]{#2}\begingroup\raggedright\endgroup

\end{document}